\documentclass[reqno,a4paper,final,oneside,11pt]{amsart}
\usepackage[utf8]{inputenc}
\usepackage{amssymb,amstext,amsthm,eucal}
\usepackage{color}
\usepackage{array}
\usepackage{geometry}
\usepackage{hyperref}
\hypersetup{%
  pdftitle   = {The return of the four- and five-dimensional preons},
  pdfkeywords = {supergravity, supersymmetry, preons, quotients, anti
    de Sitter}
  pdfauthor  = {José Figueroa-O'Farrill, Jan Gutowski and Wafic Sabra},
  pdfcreator = {\LaTeX\ with package \flqq hyperref\frqq}
}
\newcommand{\half}{\tfrac12}
\newcommand{\fso}{\mathfrak{so}}
\newcommand{\Cl}{\mathrm{C}\ell}
\newcommand{\CCl}{\mathbb{C}\ell}
\newcommand{\Spin}{\mathrm{Spin}}
\newcommand{\SO}{\mathrm{SO}}
\newcommand{\SU}{\mathrm{SU}}
\renewcommand{\O}{\mathrm{O}}
\newcommand{\RR}{\mathbb{R}}
\newcommand{\PP}{\mathbb{P}}
\newcommand{\CC}{\mathbb{C}}
\newcommand{\ZZ}{\mathbb{Z}}
\newcommand{\be}{\boldsymbol{e}}
\newcommand{\bX}{\boldsymbol{X}}
\DeclareMathOperator{\AdS}{AdS}
\DeclareMathOperator{\sech}{sech}
\renewcommand{\d}{\text{d}}
\newcommand{\ds}{\text{ds}}
\ifx\Sp\UnDeFiNeD
\newcommand{\Sp}{\mathrm{Sp}}
\else
\renewcommand{\Sp}{\mathrm{Sp}}
\fi
\ifx\1\UnDeFiNeD
\newcommand{\1}{\boldsymbol{1}}
\else
\renewcommand{\Sp}{\mathrm{Sp}}
\fi
\newcommand{\MUNCH}[1]{\relax}

\allowdisplaybreaks[1]
\begin{document}
\title[The return of the preons]{The return of the four- and
  five-dimensional preons}
\author[Figueroa-O'Farrill]{José Figueroa-O'Farrill}
\address{Maxwell Institute \& School of Mathematics, University of
  Edinburgh, UK}
\email{J.M.Figueroa@ed.ac.uk}
\author[Gutowski]{Jan Gutowski}
\address{DAMTP, Centre for Mathematical Sciences, University of
  Cambridge, UK}
\email{J.B.Gutowski@damtp.cam.ac.uk}
\author[Sabra]{Wafic Sabra}
\address{CAMS \& Physics Department, American University of Beirut,
  Lebanon}
\email{ws00@aub.edu.lb}
\date{May 2007}
\begin{abstract}
  We prove the existence of $\frac34$-BPS preons in four- and
  five-dimensional gauged supergravities by explicitly constructing
  them as smooth quotients of the $\AdS_4$ and $\AdS_5$ maximally
  supersymmetric backgrounds, respectively.  This result illustrates
  how the spacetime topology resurrects a fraction of supersymmetry
  previously ruled out by the local analysis of the Killing spinor
  equations.
\end{abstract}
\maketitle
\tableofcontents

\section{Introduction}
\label{sec:intro}

Given a $d$-dimensional supergravity theory with $n$ (real)
supercharges, it is an interesting problem to determine the existence
of backgrounds preserving the next-to-largest possible number $N_p<n$
of supersymmetries.  Depending on the vagaries of the spinor
representations and of the Killing spinor equations---i.e., whether
the space of Killing spinors is real, complex or quaternionic---the
number $N_p$ may be $n-1$, $n-2$ or $n-4$.  In any one of these cases,
the resulting background is called a supergravity (or BPS)
\textbf{preon}, generalising a notion introduced in
\cite{11dPreons,Preons} in the context of eleven-dimensional
supergravity, and recently reviewed in \cite{BAPreonsReview}.  The
name derives from the Pati--Salam model \cite{PatiSalamPreons}, and
just as the Pati--Salam preons are constituents of both quarks and
leptons, supergravity preons are meant to give rise to other BPS
backgrounds; although a precise mechanism has not been proposed.  The
argument for the existence of preons uses the correspondence between
the supersymmetry broken by a BPS state and the rank of the
``central'' charge in the relevant superalgebra.  The algebraic
manifestation of preons is simply the fact that a rank-$r$ central
charge can be written as a linear combination of $r$ rank-$1$ central
charges, whose corresponding BPS states are supposed to be the preons.

To the best of our knowledge, however, the geometric realisation in
supergravity of the addition of central charges is still an open
problem, and moreover its existence has been brought into question by
recent results ruling out supergravity preons in ten and eleven
dimensions: \cite{NoMPreons, NoIIBPreons, PreonsConfinement,
  FigGadPreons}, where $n=32$ and $N_p=31$.  Such nonexistence results
are usually obtained in two steps: in the first step one proves that a
preonic background always admits locally the maximal number of
supersymmetries, whence if it exists at all, it must be the quotient
of a maximally supersymmetric background by a discrete symmetry group.
In the second step one then uses the classification of maximally
supersymmetric solutions \cite{FOPMax} to show that no such quotients
exist.  For eleven-dimensional supergravity preons, these two steps
are described in \cite{NoMPreons} and \cite{FigGadPreons},
respectively.

The investigations in four and five dimensions flow along similar
lines.  For the case of gauged four- and five-dimensional
supergravities with matter, where $N_p/n = 3/4$, it has been shown
that $\frac34$-BPS states are locally maximally supersymmetric
backgrounds of the minimal theory, where there is no matter content
and gauge fields are flat \cite{GGS4d,GGS5d,caldarelli}.  It therefore
remains to investigate whether one can obtain $\frac34$-BPS
backgrounds by quotienting the maximally supersymmetric backgrounds,
which for the $d$-dimensional theory, are given by $\AdS_d$.

In this note, using results of \cite{FigSimAdS}, we will show that
remarkably such quotients do indeed exist.  Indeed, we will exhibit a
family (parametrised by the positive real numbers) of discrete groups
acting freely and properly discontinuously via isometries on $\AdS_d$,
preserving the spin structure, and leaving invariant a
$\frac34$-dimensional subspace of the Killing spinors.  These groups
are discrete subgroups of the one-parameter subgroup labelled (6) in
the classification of \cite[§4.1.1]{FigSimAdS}.  Such quotients have
already appeared briefly in \cite[§B.2]{FMPHom} and are also the
subject of an ongoing study \cite{JoanDLCQ} on discrete lightcone
quantisation in the AdS/CFT correspondence.

This note is organised as follows.  In §\ref{sec:quotients} we will
introduce the relevant quotients of $\AdS_d$ and prove that they are
smooth.  In fact, we will work with the hyperboloid model of $\AdS_d$,
which is already an infinite cyclic quotient by the centre, and
quotient it further by a non-central $\ZZ$-subgroup.  In
§\ref{sec:supersymmetry} we will show that such quotients preserve
$\tfrac34$ of the supersymmetry.  We will do this in two ways: one
purely representation theoretic using the Bär cone construction, and
one by explicitly calculating the Killing spinors relative to an
adapted coordinate system.  Finally in §\ref{sec:conc} we offer some
concluding remarks.

\section{A family of smooth quotients of AdS}
\label{sec:quotients}

Let $\RR^{2,d-1}$ denote the ($d+1$)-dimensional pseudo-euclidean
space with global coordinates $X_0,X_1,\dots,X_d$ and flat metric
given by
\begin{equation}
  \label{eq:flat}
  \ds^2 = -\d X_0^2 - \d X_d^2 + \sum_{i=1}^{d-1} \d X_i^2~.
\end{equation}
The hyperboloid
\begin{equation}
  \label{eq:hyper}
  X_0^2 + X_d^2 - \sum_{i=1}^{d-1} X_i^2 = \ell^2
\end{equation}
describes a lorentzian manifold with constant negative curvature.  The
scalar curvature is readily calculated to be
\begin{equation}
  \label{eq:curv}
  R = -\frac{d(d-1)}{\ell^2}~,
\end{equation}
whence it is locally isometric to $\AdS_d$ with radius of curvature
$\ell$.  It is in fact the quotient of $\AdS_d$ by the action of the
centre of the isometry group and will be hereafter referred to,
somewhat loosely, as the \textbf{hyperboloid model} for $\AdS_d$.  The
virtue of this model is that the isometries of $\AdS_d$ are realised
linearly in the embedding space, namely as $\O(2,d-1)$ acting linearly
on $\RR^{2,d-1}$.

Let $k>0$ and consider the element $\gamma \in \SO(2,d-1)$
defined by
\begin{equation}
  \label{eq:gamma}
  \begin{pmatrix}
    X_d \\ X_0 \\ X_1 \\ X_2 \\ X_i
  \end{pmatrix} \mapsto
  \begin{pmatrix}
    X_d + k (X_2 - X_0)\\
    X_0 + k (X_d - X_1)\\
    X_1 + k (X_2 - X_0)\\
    X_2 + k (X_d - X_1)\\
    X_i
  \end{pmatrix}~,
\end{equation}
for $2<i<d$.  The element $\gamma$ is in the image of the
exponential map
\begin{equation}
  \label{eq:generator}
  \gamma = \exp(k Z)~,
\end{equation}
where $Z \in \fso(2,d-1)$ is given by
\begin{equation}
  \label{eq:liealg}
  Z = (\d X_0 - \d X_2) \wedge (\d X_1 - \d X_d)~,
\end{equation}
where we have identified $\fso(2,d-1)$ with the constant-coefficient
$2$-forms on $\RR^{2,d-1}$.  For any $k>0$, $\gamma$ generates an
infinite cyclic subgroup $G_k < \SO(2,d-1)$, acting freely on the
hyperboloid.  Indeed the fixed points of $\gamma$ on $\RR^{2,d-1}$ lie
on the ($d-1$)-dimensional plane cut out by the equations $X_0 = X_2$
and $X_1 = X_d$, but this plane does not intersect the hyperboloid.

The group $G_k$ also acts properly discontinuously, as we now show.
It is enough to show that for every point $p$ and every compact set
$K$ in the hyperboloid, the intersection $G_k \cdot p \cap K$ of $K$
and the orbit of $p$ has finite cardinality.  The hyperboloid inherits
the subspace topology from $\RR^{2,d-1}$, which is the standard metric
topology induced from the \emph{euclidean} norm in $\RR^{d+1}$, whence
a compact set $K$ on the hyperboloid is the intersection of the
hyperboloid with a closed and bounded subset of $\RR^{d+1}$.  It is
therefore enough to show that any ball of finite radius centred at $p$
contains finitely many points of the orbit $G_k \cdot p$.  To see
this, let $p$ have coordinates $\bX = (X_0,X_1,\dots,X_d)$ and
consider the euclidean distance between $p$ and the point $\gamma^N
\cdot p$ in its $G_k$-orbit, which is given by
\begin{equation}
  \|\gamma^N \bX - \bX \|^2 = 2 N^2 k^2 \left( (X_2-X_0)^2 +
    (X_d-X_1)^2 \right)~.
\end{equation}
Although the expression in parenthesis can be arbitrarily small on the
hyperboloid, it is always positive, hence for any $k > 0$ and any
finite radius $L$, there will be some $N_0$, depending on the point
$p$, for which
\begin{equation}
  \|\gamma^N \bX - \bX \| > L \qquad \text{for all $N>N_0$.}
\end{equation}
Therefore at most $N_0$ points in the orbit will lie inside the ball.

In summary, $G_k$ acts freely and properly discontinuously on the
hyperboloid and hence the quotient is smooth and locally isometric to
$\AdS_d$.   As shown in \cite[§2.4]{FOMRS} a $\ZZ$-quotient always
preserves the spin structure, hence we can ask how much supersymmetry
such a quotient preserves.  In the next section we answer this
question and show that it preserves $\frac34$ of the supersymmetry of
$\AdS_d$, for $d=4,5$.

\section{Supersymmetry}
\label{sec:supersymmetry}

As shown in \cite{GGS4d,GGS5d}, the supersymmetries of the maximally
supersymmetric backgrounds of four- and five-dimensional gauged
supergravities are in one-to-one correspondence with \textbf{geometric
  Killing spinors}; that is, spinor fields $\varepsilon$
obeying\footnote{The factor of $i$ is due to our Clifford algebra
  conventions: $\Gamma_A\Gamma_B + \Gamma_B\Gamma_A = + 2 \eta_{AB}$,
  with $\eta$ mostly plus in lorentzian signature.}
\begin{equation}
  \label{eq:Killing}
  \nabla_a \varepsilon = i \lambda \Gamma_a \varepsilon~,
\end{equation}
where the \textbf{Killing constant} $\lambda$ is related to the scalar
curvature by
\begin{equation}
  \label{eq:curvature}
  R = 4\lambda^2 d(d-1)~,
\end{equation}
whence it can be either real or pure imaginary.  Maximal supersymmetry
is attained by the spaces of constant curvature and for these theories
it is $\AdS_d$, for which $\lambda$ is pure imaginary.  In either of these
theories, preons will exist if there exists some discrete subgroup of
isometries of $\AdS_d$, for $d=4,5$, leaving invariant a
$\frac34$-dimensional subspace of Killing spinors.

In this section we will present two proofs that the quotients
described above of $\AdS_d$, for $d=4,5$, do indeed preserve
$\tfrac34$ of the supersymmetry.  The first proof uses representation
theory and the cone construction and is computationally very simple.
On the other hand, the second proof is more elementary but
computationally more involved and follows from an explicit calculation
of the Killing spinors relative to a coordinate system adapted to the
group action.

\subsection{The cone construction and representation theory}
\label{sec:cone}

As discussed above, the supersymmetries in the $\AdS_d$ vacuum are in
one-to-one correspondence with the geometric Killing spinors, and
those of the quotient by a group $G$ (assumed to preserve the spin
structure) are precisely the $G$-invariant Killing spinors.  In this
section we will show how to reinterpret this in terms of
representation theory.

The main technical tool is the cone construction \cite{Baer,KathHabil}
relating geometric Killing spinors \eqref{eq:Killing} on a manifold
$(M,g)$ to parallel spinors on its cone, a manifold one dimension
higher with metric $dr^2 + 4 \lambda^2 r^2 g$.  If $g$ is riemannian
and $\lambda$ real, the cone metric is again riemannian.  We are
interested however in the case of $g$ lorentzian and $\lambda$ pure
imaginary, whence the cone has signature $(2,d-1)$.  Since $\AdS_d$
has constant curvature, its cone is an open subset of the
pseudo-euclidean space $\RR^{2,d-1}$ with the standard flat metric,
corresponding to those vectors with negative norm.

Relative to flat coordinates for $\RR^{2,d-1}$ and the corresponding
global orthonormal frame, parallel spinors are in one-to-one
correspondence with \emph{constant} $\Spin(2,d-1)$ spinors.  In the
case of odd $d$, the spinors are chiral: the choice of chirality
corresponding to a choice of sign in the Killing constant $\lambda$ in
equation \eqref{eq:Killing}.  For even $d$, there is up to equivalence
a unique spinor representation, and the sign of the Killing constant
enters in the choice of embedding $\Cl(d-1,1) \subset \Cl(2,d-1)$
between Clifford algebras.  This is explained in detail in
\cite{Baer,KathHabil} and in a supergravity context in the forthcoming
paper \cite{FigLeiSim}.

The isometry group of the simply-connected $\AdS_d$ is the universal
covering group $\widetilde{\SO}(2,d-1)$ of $\SO(2,d-1)$.  As
discussed, for example, in \cite{FigSimAdS,FOMRS}, the action of the
spin cover $\widetilde{\Spin}(2,d-1)$ on the Killing spinors is not
effective, but factors through the action of $\Spin(2,d-1)$, which is
an infinite cyclic quotient.  Let $\widetilde G <
\widetilde{\SO}(2,d-1)$ be a discrete subgroup acting freely and
properly discontinuously on $\AdS_d$.  The resulting quotient
$\AdS_d/\widetilde G$ is smooth and moreover will be spin if
$\widetilde G$ lifts isomorphically to a subgroup of
$\widetilde{\Spin}(2,d-1)$, which we also denote $\widetilde G$.
Assume this is so and let $G < \Spin(2,d-1)$ denote its projection
onto $\Spin(2,d-1)$.  Then the Killing spinors on the quotient
$\AdS_d/\widetilde G$ are precisely the $G$-invariant Killing spinors
on $\AdS_d$.  Now, as discussed in \cite{JMFKilling}, the cone
construction is equivariant under the action of isometries.  Therefore
the $G$-invariant Killing spinors on $\AdS_d$ are precisely the
$G$-invariant parallel spinors on $\RR^{2,d-1}$; in other words, the
$G$-invariant subspace in the (chiral) spinor representation of
$\Spin(2,d-1)$.

The groups $G_k$ in question are in the image of the exponential map,
hence their image in the Spin groups lie in the identity component.
For $\AdS_4$, this is the identity component of $\Spin(2,3)$ which is
isomorphic to $\Sp(4,\RR)$.  The spinor representation is therefore
real and 4-dimensional, and we will show that $G_k$ fixes a
3-dimensional subspace.  For $\AdS_5$, on the other hand, it is the
identity component of $\Spin(2,4)$, which is isomorphic to $\SU(2,2)$.
The relevant spinorial representation is therefore complex and
4-dimensional, and we will show that $G_k$ fixes a 3-dimensional
complex subspace.  The calculation follows \emph{mutatis mutandis} the
approach of \cite[§6.1.4]{FigSimAdS}.

In fact, we will work with any $d>2$.  As usual when working with the
spin groups, it is convenient to embed them in the Clifford algebra.
The image $\widehat\gamma \in \Spin(2,d-1)$ of the generator $\gamma$
in \eqref{eq:generator} is given by the exponential
\begin{equation}
  \widehat\gamma = \exp \left(\half k
    (\Gamma_0-\Gamma_2)(\Gamma_1-\Gamma_d)\right) \in \Spin(2,d-1)
  \subset \Cl(2,d-1)
\end{equation}
of the Lie algebra element \eqref{eq:liealg}.  Using the fact that
both $\Gamma_0-\Gamma_2$ and $\Gamma_1-\Gamma_d$ anticommute and
square to zero, we see that
\begin{equation}
  \widehat\gamma = 1 + \half k
  (\Gamma_0-\Gamma_2)(\Gamma_1-\Gamma_d)~.
\end{equation}

Let $V$ denote the relevant irreducible Clifford representation of
$\Cl(2,d-1)$.  For even $d$ there are actually two inequivalent
representations, but the cone construction chooses one of them
\emph{ab initio}.  Let us introduce four subspaces of $V$ given by
\begin{equation}
  V_{\pm\pm} = \ker (\Gamma_0\pm\Gamma_2) \cap \ker (\Gamma_1\pm
  \Gamma_d)
\end{equation}
with uncorrelated signs.  It is plain to see that
\begin{equation}
  V = V_{++} \oplus V_{+-} \oplus V_{-+} \oplus V_{--}
\end{equation}
and that each subspace is $\frac14$-dimensional.  Relative to this
decomposition, the matrix of $\widehat\gamma$ is given by
\begin{equation}
  \widehat\gamma =
  \begin{pmatrix}
    1 & 0 & 0 & 0\\
    0 & 1 & 0 & 0 \\
    0 & 0 & 1 & 0 \\
    \Phi & 0 & 0 & 1
  \end{pmatrix}~,
\end{equation}
where $\Phi := \half k (\Gamma_0-\Gamma_2)(\Gamma_1-\Gamma_d):
V_{++} \stackrel{\cong}{\to} V_{--}$ has no kernel.  Therefore there
is a $\frac34$-dimensional invariant subspace given by
\begin{equation}
  V^{G_k} = V_{+-} \oplus V_{-+} \oplus V_{--}~.
\end{equation}

When $d$ is odd, we have to restrict further to chiral spinors, but it
is clear that the above discussion still holds if we substitute for
$V$ the relevant chiral spinor representation from the start.  In
either case, $G_k$ preserves a $\frac34$-dimensional subspace of the
relevant spinorial representation.

\subsection{Explicit construction of the Killing spinors}
\label{sec:spin}

We will now explicitly solve for the Killing spinors on the
hyperboloid relative to a coordinate system adapted to the action of
the group $G_k$.  The calculation is aided by representing spinors as
exterior forms as described, for example, in \cite{Wang,LM,Harvey}.
Although the discussion can be made general, we will focus for
definiteness on the four- and five-dimensional cases of interest.

Let $\Cl(2,2)$ be the Clifford algebra of the pseudo-euclidean space
$\RR^{2,2}$.  It is well-known that the unique irreducible
representation of $\Cl(2,2)$ is isomorphic to the exterior algebra
$\Lambda P$, for $P\subset\RR^{2,2}$ any isotropic plane.  More
concretely, one builds the isomorphism as follows.  Given $P$ one
chooses a complementary isotropic plane $P'$, so that $\RR^{2,2} = P
\oplus P'$.  Under the inner product, $P'$ is naturally isomorphic to
the dual $P^*$ of $P$.  The action of the Clifford algebra on $\Lambda
P$ is uniquely defined by declaring gamma matrices in $P$ to act via
the wedge product and those in $P'$ to act by the contraction with the
corresponding element of $P^*$.  Moreover as the Clifford group acts
transitively on the space of isotropic planes, all choices of $P$
yield equivalent representations.

In this note, however, we are interested in $\Cl(1,3)$, and in
$\RR^{1,3}$ there are no isotropic planes.  Nevertheless we can still
identify spinors of $\Cl(1,3)$ with exterior forms at the price of
complexifying the Clifford algebra.  We complexify $\RR^{1,3}$ to
$\CC^4$ extending the inner product to a complex bilinear form.  The
resulting complex Clifford algebra is denoted $\CCl(4)$, where we no
longer keep track of the signature because there is no such notion for
a complex inner product.  We may now choose complementary isotropic
(complex) planes $\CC^4 = \PP \oplus \PP'$ and define the action of
$\CCl(4)$ on $\Lambda \PP$ as we did before.  Again one shows that up to
equivalence this is independent of the choice of isotropic plane.  To
recover the action of $\Cl(1,3)$ we restrict this representation of
$\CCl(4)$ to a suitable real section of $\CC^4$ on which the complex
inner product has signature $(1,3)$.

Concretely, if we let $\be^1,\be^2$ span an isotropic plane $\PP$ and
let $\be_1,\be_2$ be the canonical dual basis for $\PP^*$, then the
following linear transformations of $\Lambda \PP$
\begin{equation}
  \label{eq:exterior}
  \begin{aligned}[m]
  \Gamma_0 &:= - \be^2 \wedge {} + {} \imath_{\be_2}\\
  \Gamma_2 &:= \be^2 \wedge {} + {} \imath_{\be_2}
  \end{aligned}
  \qquad\qquad
  \begin{aligned}[m]
  \Gamma_1 &:= \be^1 \wedge {} + {} \imath_{\be_1}\\
  \Gamma_3 &:= i \be^1 \wedge {} - {} i \imath_{\be_1}
  \end{aligned}
\end{equation}
make $\Lambda \PP$ into an irreducible representation of $\Cl(1,3)$.
We may also extend this to an irreducible representation of
$\Cl(1,4)$ on the same space by defining $\Gamma_4 := \pm i
\Gamma_0\Gamma_1\Gamma_2\Gamma_3$.  The choice of sign merely reflects
the fact that $\Cl(1,4)$ has two inequivalent irreducible
representations.

There are two spin structures on the hyperboloid models for $\AdS_4$
and $\AdS_5$, due to the noncontractible timelike circles $X_0^2 +
X_d^2 = \text{constant}$.  Choosing the trivial spin structure, we may
trivialise the spinor bundle and effectively represent a spinor
$\varepsilon$ as a function from the hyperboloid to the exterior
algebra $\Lambda\CC^2$ generated by $\be^1$ and $\be^2$:
\begin{equation}
  \label{eq:spinorasform}
  \varepsilon = f_0 \, \1 + f_1 \be^1 + f_2 \be^2 + f_3 \be^{12}~,
\end{equation}
where $\be^{12} = \be^1 \wedge \be^2$ and the $f_i$, for $i=0,1,2,3$
are complex-valued functions.

\subsubsection{Killing spinors for $\AdS_4$}
\label{sec:ads4}

We start by considering the case of $\AdS_4$.  Let us parametrise the
hyperboloid with coordinates $t, x, r, \rho$ and set
\begin{equation}
  \begin{aligned}[m]
    X_4 &= \frac{\ell }{2} \cosh \rho \left((r+r^{-1}) \cos t
      - r x \sin t \right) \\
    X_0 &= \frac{\ell}{2} \cosh \rho \left( (r+r^{-1}) \sin t + r x
      \cos t \right)\\
    X_1 &= \frac{\ell }{2} \cosh \rho \left((-r+r^{-1}) \cos t - r x
      \sin t \right)\\
    X_2 &= \frac{\ell }{2} \cosh \rho \left((-r+r^{-1}) \sin t + r x
      \cos t \right)\\
    X_3 &= \ell \sinh \rho~,
  \end{aligned}
\end{equation}
for $x,\rho \in \RR$, $t \in [0,2\pi)$ and $r>0$.  In these
coordinates, the action of $\gamma$ consists in shifting the $x$
coordinate: $x \mapsto x + 2k$.

On pulling back the metric of $\RR^{2,3}$ to the hyperboloid we find
the following metric
\begin{equation}
  \ell^{-2} \ds^2 = \d\rho^2 + \cosh^2 \rho \left( - (\d t+\half r^2 \d
    x)^2 + \tfrac14 r^4 \d x^2 + r^{-2} \d r^2 \right)~.
\end{equation}
We therefore take the following orthonormal coframe
\begin{equation}
  \begin{aligned}[b]
    \theta^0 &= \ell \cosh \rho \left(\d t + \half r^2 \d x\right)\\
    \theta^2 &= \ell \d\rho
  \end{aligned}
  \qquad\qquad
  \begin{aligned}[b]
    \theta^1 &= \half \ell r^2 \cosh \rho \d x\\
    \theta^3 &= \ell r^{-1} \cosh \rho \d r~.
  \end{aligned}
\end{equation}
In this basis, the nonvanishing components of the spin connection
are given by
\begin{equation}
  \begin{aligned}[m]
    \omega_0{}^{02} &= \ell^{-1} \tanh \rho\\
    \omega_1{}^{03} &= \ell^{-1}\sech \rho\\
    \omega_1{}^{13} &= 2\ell^{-1}\sech \rho\\
    \omega_3{}^{23} &= - \ell^{-1} \tanh \rho
  \end{aligned}
  \qquad\qquad
  \begin{aligned}[m]
    \omega_0{}^{13} &= - \ell^{-1}\sech \rho\\
    \omega_1{}^{12} &= \ell^{-1} \tanh \rho\\
    \omega_3{}^{01} &= -\ell^{-1}\sech \rho~.\\
  \end{aligned}
\end{equation}
The Killing spinor equation is given by
\begin{equation}
  \label{eq:KillingSpinor}
  \left(\partial_\mu + \tfrac14 \omega_\mu{}^{\nu_1\nu_2}
    \Gamma_{\nu_1\nu_2} + \half \ell^{-1} \Gamma_\mu\right)
  \varepsilon = 0~,
\end{equation}
and it is straightforward but tedious to show that the space of
Killing spinors is spanned over $\CC$ by $\varepsilon_i$, for
$i=1,\dots,4$, where
\begin{equation}
  \label{eq:KS4}
  \begin{aligned}[m]
    \varepsilon_1 &= 2 r \left( \cosh\tfrac\rho2 - i \sinh\tfrac\rho2
    \right) (\1+\be^{12}) + 2 r \left( \sinh\tfrac\rho2 - i
      \cosh\tfrac\rho2 \right) (\be^1-\be^2)\\
    \varepsilon_2 &= 2 e^{it} \left( \cosh\tfrac\rho2 + i
      \sinh\tfrac\rho2\right) \, \1 - 2 e^{it} \left( \sinh\tfrac\rho2
      + i
      \cosh\tfrac\rho2 \right) \be^2\\
    \varepsilon_3 &= 2 e^{-it} \left( \cosh\tfrac\rho2 - i
      \sinh\tfrac\rho2\right) \be^1 + 2 e^{-it} \left(\sinh\tfrac\rho2
      - i
      \cosh\tfrac\rho2 \right) \be^{12}\\
    \varepsilon_4 &= \tfrac2r (1 - i r^2 x) \left( \cosh\tfrac\rho2 -
      i \sinh\tfrac\rho2 \right) \, \1 - \tfrac2r (1 + i r^2 x)
    \left(\sinh\tfrac\rho2 - i \cosh\tfrac\rho2 \right)
    \be^1\\
    & \quad {} - \tfrac2r (1-i r^2 x) \left( \sinh\tfrac\rho2 - i
      \cosh\tfrac\rho2\right) \be^2 - \tfrac2r (1 + i r^2 x) \left(
      \cosh\tfrac\rho2 - i \sinh\tfrac\rho2 \right)\be^{12}~.
  \end{aligned}
\end{equation}
Note that although $\varepsilon_4$ depends linearly on $x$, the other
three basis elements are independent of $x$.  Hence, under the
identification $x \sim x+2k$, $\varepsilon_i$, for $i=1,2,3$, remain
globally well-defined whereas $\varepsilon_4$ does not, resulting in
one fourth of the supersymmetry being broken.

\subsubsection{Killing spinors for $\AdS_5$}
\label{sec:ads5}

The calculation for $\AdS_5$ is very similar.  We parametrise the
hyperboloid now by coordinates $t,x,\rho,r,\theta$ as follows
\begin{equation}
  \begin{aligned}[m]
    X_5 &= \frac{\ell }{2} \cosh \rho \left((r+r^{-1}) \cos t
      - r x \sin t \right) \\
    X_0 &= \frac{\ell}{2} \cosh \rho \left( (r+r^{-1}) \sin t + r x
      \cos t \right)\\
    X_1 &= \frac{\ell }{2} \cosh \rho \left((-r+r^{-1}) \cos t - r x
      \sin t \right)\\
    X_2 &= \frac{\ell }{2} \cosh \rho \left((-r+r^{-1}) \sin t + r x
      \cos t \right)\\
    X_3 &= \ell \sinh \rho \cos\varphi\\
    X_4 &= \ell \sinh \rho \sin\varphi~,
  \end{aligned}
\end{equation}
where now $x\in \RR$, $t,\varphi \in [0,2\pi)$ and $\rho,r>0$.
Because of the condition $\rho>0$ this chart does not cover all of the
hyperboloid: there is a codimension-2 hyperboloid missing.  In these
coordinates, the action of $\gamma$ again consists in shifting the $x$
coordinate: $x \mapsto x + 2k$.

The induced metric on the hyperboloid is now given by
\begin{equation}
  \ell^{-2} \ds^2 = \d\rho^2 + \cosh^2 \rho \left( - (\d t+\half r^2 \d
    x)^2 + \tfrac14 r^4 \d x^2 + r^{-2} \d r^2 \right) + \sinh^2\rho
  \d\varphi^2~.
\end{equation}
We therefore take the following orthonormal coframe
\begin{equation}
  \begin{aligned}[m]
    \theta^0 &= \ell \cosh \rho \left(\d t + \half r^2 \d x\right)\\
    \theta^2 &= \ell \d\rho\\
    \theta^4 &= \ell \sinh\rho \d \varphi
  \end{aligned}
  \qquad\qquad
  \begin{aligned}[m]
    \theta^1 &= \half \ell r^2 \cosh \rho \d x\\
    \theta^3 &= \ell r^{-1} \cosh \rho \d r~.
  \end{aligned}
\end{equation}
In this basis, the nonvanishing components of the spin connection
are given by
\begin{equation}
  \begin{aligned}[m]
    \omega_0{}^{02} &= \ell^{-1} \tanh \rho\\
    \omega_1{}^{03} &= \ell^{-1}\sech \rho\\
    \omega_1{}^{13} &= 2\ell^{-1}\sech \rho\\
    \omega_3{}^{23} &= - \ell^{-1} \tanh \rho
  \end{aligned}
  \qquad\qquad
  \begin{aligned}[m]
    \omega_0{}^{13} &= - \ell^{-1}\sech \rho\\
    \omega_1{}^{12} &= \ell^{-1} \tanh \rho\\
    \omega_3{}^{01} &= -\ell^{-1}\sech \rho\\
    \omega_4{}^{24} &= -\ell^{-1}\coth \rho~,
  \end{aligned}
\end{equation}
whereas the Killing spinor equation is again given by
\eqref{eq:KillingSpinor}.  The form of the connection is such that the
Killing spinor equations for $\AdS_5$ consist of the Killing spinor
equations for $\AdS_4$ and an extra equation involving the new
coordinate $\varphi$.  Hence Killing spinors of $\AdS_5$ have the
form
\begin{equation}
  \psi = \sum_{i=1}^4 a_i(\varphi) \varepsilon_i~,
\end{equation}
where the $\varepsilon_i$ are given in \eqref{eq:KS4}, and are subject
to the equation
\begin{equation}
  \left(\frac{\partial}{\partial\varphi} + \half
  \omega_{\varphi}^{24}\Gamma_{24} + \half \ell^{-1}
  \Gamma_{\varphi}\right) \psi = 0~,
\end{equation}
which merely fixes the $\varphi$-dependence of the functions
$a_i(\varphi)$.  In summary, the space of Killing spinors for
$\AdS_5$ is spanned over $\CC$ by $\psi_i$, for $i=1,2,3,4$, given
by
\begin{equation}
    \psi_1 = e^{\mp i\varphi/2} \varepsilon_1~,\quad
    \psi_2 = e^{\pm i\varphi/2} \varepsilon_2~, \quad
    \psi_3 = e^{\pm i\varphi/2} \varepsilon_3~,\quad\text{and}\quad
    \psi_4 = e^{\mp i\varphi/2} \varepsilon_4~,
\end{equation}
where the sign corresponds to the choice of irreducible representation
of $\Cl(1,4)$ and where the $\varepsilon_i$ are again given in
\eqref{eq:KS4}.

It would appear that the $\psi_i$ are not globally well-defined, as
they are antiperiodic when one takes $\varphi \mapsto \varphi+2
\pi$. However, this is merely an artifact of the choice of basis.  To
see this, make the $\Spin(1,4)$ gauge transformation
\begin{equation}
  \psi_i \mapsto \psi_i' = e^{-{\varphi \over 2} \Gamma_{24}} \psi_i~,
\end{equation}
which induces a rotation by angle $\varphi$ in the $(24)$
plane at the level of the coframes.  Under this gauge transformation,
the Killing spinors are now $2\pi$-periodic in $\varphi$
and, as a bonus, have a well-defined limit as $\rho \to 0$:
\begin{equation}
  \label{eq:KS5}
  \begin{aligned}[m]
    \psi_1' &= 2 r \left(\cosh\tfrac\rho2 - i e^{\mp i \varphi}
      \sinh\tfrac\rho2 \right) (\1+\be^{12}) + 2 r \left(e^{\mp i \varphi}\sinh\tfrac\rho2 - i
      \cosh\tfrac\rho2 \right) (\be^1-\be^2)\\
    \psi_2' &= 2 e^{it} \left(
      \cosh\tfrac\rho2 + i e^{\pm i \varphi}\sinh\tfrac\rho2\right) \, \1 - 2 e^{it} \left(
      e^{\pm i \varphi}\sinh\tfrac\rho2
      + i\cosh\tfrac\rho2 \right) \be^2\\
    \psi_3' &= 2 e^{-it}\left(
      \cosh\tfrac\rho2 - ie^{\pm i \varphi}\sinh\tfrac\rho2\right) \be^1 + 2 e^{-it}
    \left(e^{\pm i \varphi}\sinh\tfrac\rho2 - i \cosh\tfrac\rho2 \right) \be^{12}\\
    \psi_4' &= \tfrac2r (1 - i r^2 x) \left(
      \cosh\tfrac\rho2 - i e^{\mp i \varphi}\sinh\tfrac\rho2 \right) \, \1 - \tfrac2r (1 + i r^2 x)
      \left(e^{\mp i \varphi}\sinh\tfrac\rho2 - i\cosh\tfrac\rho2 \right) \be^1\\
    & \quad {} - \tfrac2r (1-i r^2 x)
    \left(e^{\mp i \varphi}\sinh\tfrac\rho2 - i \cosh\tfrac\rho2\right) \be^2 - \tfrac2r (1 + i r^2 x)
    \left( \cosh\tfrac\rho2 - ie^{\mp i \varphi}\sinh\tfrac\rho2 \right)\be^{12}~.
  \end{aligned}
\end{equation}
Again we notice that whereas $\psi'_i$, for $i=1,2,3$, do not depend
on $x$, $\psi'_4$ depends linearly on $x$, whence it is not invariant
under shifting $x$.  As before, we conclude that precisely one fourth
of the supersymmetry is broken.

\section{Conclusion}
\label{sec:conc}

In this note we have shown that, contrary to expectations, there exist
backgrounds of gauged four- and five-dimensional supergravities which
are $\frac34$-BPS.  This was done by exhibiting a family of
$\ZZ$-quotients of the hyperboloid model for $\AdS_d$ preserving a
$\frac34$-dimensional  subspace of the Killing spinors.  In terms of
the simply-connected $\AdS_d$, these constitute a family of ($\ZZ
\oplus \ZZ$)-quotients, parametrised by the positive real numbers.

This result was unexpected because, as shown in \cite{GGS4d,GGS5d},
such backgrounds are locally maximally supersymmetric, and the similar
result in ten- and eleven-dimensional supergravities
\cite{NoMPreons,NoIIBPreons} was shown to survive even when the
spacetime is not simply connected \cite{FigGadPreons}.  In the cases
studied here, however, the supersymmetry is broken by global effects
and to our knowledge these are the first known cases where a fraction
of supersymmetry previously discarded by a local analysis of the
Killing spinor equations is resurrected by the topology of the
underlying spacetime.  It should also sound a note of caution
concerning the use of the holonomy algebra to rule out supersymmetry
fractions.

\section*{Acknowledgments}

We are grateful to an anonymous referee for pointing out an
imprecision in an earlier version of this paper.  In addition, JMF
would like to thank Joan Simón for sending him his notes
\cite{JoanDLCQ} prior to publication as well as for the collaboration
leading up to \cite{FigSimAdS}, which continues to prove surprisingly
useful. The work of W.~Sabra was supported in part by the National
Science Foundation under grant number PHY-0601213.

\bibliographystyle{utphys}
\bibliography{AdS,AdS3,ESYM,Sugra,Geometry}

\providecommand{\href}[2]{#2}\begingroup\raggedright\begin{thebibliography}{10}

\bibitem{11dPreons}
I.~A. Bandos, J.~A. de~Azcárraga, J.~M. Izquierdo, and J.~Lukierski, ``{BPS}
  states in {M}-theory and twistorial constituents,'' {\em Phys. Rev. Lett.}
  {\bf 86} (2001) 4451--4454, \href{http://arXiv.org/abs/hep-th/0101113}{{\tt
  hep-th/0101113}}.

\bibitem{Preons}
I.~A. Bandos, J.~A. de~Azcárraga, J.~M. Izquierdo, M.~Picón, and O.~Varela,
  ``On {BPS} preons, generalized holonomies and ${D}{=}11$ supergravities,''
  {\em Phys. Rev.} {\bf D69} (2004) 105010,
  \href{http://arXiv.org/abs/hep-th/0312266}{{\tt hep-th/0312266}}.

\bibitem{BAPreonsReview}
I.~A. Bandos and J.~A. de~Azcárraga, ``{BPS} preons in {M}-theory and
  supergravity,'' {\em Fortschr. Phys.} {\bf 55} (2007), no.~5--7, 692--698,
  \href{http://arXiv.org/abs/hep-th/0702099}{{\tt hep-th/0702099}}.

\bibitem{PatiSalamPreons}
J.~C. Pati and A.~Salam, ``Lepton number as the fourth ``color'','' {\em Phys.
  Rev. D} {\bf 10} (1974), no.~1, 275--289.

\bibitem{NoMPreons}
U.~Gran, J.~Gutowski, G.~Papadopoulos, and D.~Roest, ``{$N = 31$}, {$D =
  11$},'' \href{http://arXiv.org/abs/hep-th/0610331}{{\tt hep-th/0610331}}.

\bibitem{NoIIBPreons}
U.~Gran, J.~Gutowski, G.~Papadopoulos, and D.~Roest, ``{$N = 31$} is not
  {IIB},'' \href{http://arXiv.org/abs/hep-th/0606049}{{\tt hep-th/0606049}}.

\bibitem{PreonsConfinement}
I.~A. Bandos, J.~A. de~Azcárraga, and O.~Varela, ``On the absence of {BPS}
  preonic solutions in {IIA} and {IIB} supergravities,'' {\em J. High Energy
  Phys.} {\bf 09} (2006) 009, \href{http://arXiv.org/abs/hep-th/0607060}{{\tt
  hep-th/0607060}}.

\bibitem{FigGadPreons}
J.~M. Figueroa-O'Farrill and S.~Gadhia, ``M-theory preons cannot arise by
  quotients,'' {\em J. High Energy Phys.} {\bf 06} (2007) 043,
  \href{http://arXiv.org/abs/hep-th/0702055}{{\tt hep-th/0702055}}.

\bibitem{FOPMax}
J.~M. Figueroa-O'Farrill and G.~Papadopoulos, ``Maximal supersymmetric
  solutions of ten- and eleven-dimensional supergravity,'' {\em J. High Energy
  Phys.} {\bf 03} (2003) 048, \href{http://arXiv.org/abs/hep-th/0211089}{{\tt
  hep-th/0211089}}.

\bibitem{GGS4d}
J.~Grover, J.~Gutowski, and W.~Sabra, ``Maximally minimal preons in four
  dimensions,'' \href{http://arXiv.org/abs/hep-th/0610128}{{\tt
  hep-th/0610128}}.

\bibitem{GGS5d}
J.~Grover, J.~Gutowski, and W.~Sabra, ``Vanishing preons in the fifth
  dimension,'' {\em Class. Quant. Grav.} {\bf 24} (2007) 417--432,
  \href{http://arXiv.org/abs/hep-th/0608187}{{\tt hep-th/0608187}}.

\bibitem{caldarelli}
S.~L. Cacciatori, M.~M. Caldarelli, D.~Klemm, and D.~S. Mansi, ``More on {BPS}
  solutions of {$N{=}2$}, {$d=4$} gauged supergravity,'' {\em JHEP} {\bf 07}
  (2004) 061, \href{http://arXiv.org/abs/hep-th/0406238}{{\tt hep-th/0406238}}.

\bibitem{FigSimAdS}
J.~M. Figueroa-O'Farrill and J.~Simón, ``Supersymmetric {K}aluza--{K}lein
  reductions of {AdS} backgrounds,'' {\em Adv. Theor. Math. Phys.} {\bf 8}
  (2004) 217--317, \href{http://arXiv.org/abs/hep-th/0401206}{{\tt
  hep-th/0401206}}.

\bibitem{FMPHom}
J.~M. Figueroa-O'Farrill, P.~Meessen, and S.~Philip, ``Supersymmetry and
  homogeneity of {M}-theory backgrounds,'' {\em Class. Quant. Grav.} {\bf 22}
  (2005) 207--226, \href{http://arXiv.org/abs/hep-th/0409170}{{\tt
  hep-th/0409170}}.

\bibitem{JoanDLCQ}
J.~Simón, ``Notes on {DLCQ} in {AdS/CFT}.'' Private communication.

\bibitem{FOMRS}
J.~M. Figueroa-O'Farrill, O.~Madden, S.~Ross, and J.~Simón, ``Quotients of
  {$\AdS_{p+1}\times S^q$}: causally well-behaved spaces and black holes,''
  {\em Phys. Rev.} {\bf D69} (2004) 124026,
  \href{http://arXiv.org/abs/hep-th/0402094}{{\tt hep-th/0402094}}.

\bibitem{Baer}
C.~B{\"a}r, ``Real {K}illing spinors and holonomy,'' {\em Comm. Math. Phys.}
  {\bf 154} (1993) 509--521.

\bibitem{KathHabil}
I.~Kath, ``Killing spinors on pseudo-riemannian manifolds,'' 1999.
\newblock Habilitationsschrift, Humboldt-Universit\"at zu Berlin.

\bibitem{FigLeiSim}
J.~M. Figueroa-O'Farrill, F.~Leitner, and J.~Sim\'on, ``Supersymmetric
  {F}reund--{R}ubin backgrounds.'' In preparation.

\bibitem{JMFKilling}
J.~M. Figueroa-O'Farrill, ``On the supersymmetries of {A}nti-de~{S}itter
  vacua,'' {\em Class. Quant. Grav.} {\bf 16} (1999) 2043--2055,
  \href{http://arXiv.org/abs/hep-th/9902066}{{\tt hep-th/9902066}}.

\bibitem{Wang}
M.~Wang, ``Parallel spinors and parallel forms,'' {\em Ann. Global Anal. Geom.}
  {\bf 7} (1989), no.~1, 59--68.

\bibitem{LM}
H.~Lawson and M.~Michelsohn, {\em Spin geometry}.
\newblock Princeton University Press, 1989.

\bibitem{Harvey}
F.~Harvey, {\em Spinors and calibrations}.
\newblock Academic Press, 1990.

\end{thebibliography}\endgroup

\end{document}